\documentclass[11pt]{article}
\usepackage{fullpage,ogonek}

\newtheorem{theorem}{Theorem}
\newtheorem{lemma}[theorem]{Lemma}

\newcommand{\polylog}{\ensuremath{\mathrm{polylog}}}

\begin{document}

\title{Space-efficient RLZ-to-LZ77 conversion}
\author{Travis Gagie}
\maketitle

\begin{abstract}
Consider a text $T [1..n]$ prefixed by a reference sequence $R = T [1..\ell]$.  We show how, given $R$ and the $z'$-phrase relative Lempel-Ziv parse of $T [\ell + 1..n]$ with respect to $R$, we can build the LZ77 parse of $T$ in $n\,\polylog (n)$ time and $O (\ell + z')$ total space.
\end{abstract}

\section{Introduction}

Ziv and Lempel's~\cite{ZL77} LZ77 is a classic compression scheme still widely used today, achieving excellent compression particularly on highly repetitive datasets such as massive genomic databases.  Building the LZ77 parse for such a database in a reasonable amount of workspace is challenging, however, so researchers have developed easy-to-build variants that achieve comparable compression in practice.  As an implementation of Ziv-Merhav cross parsing~\cite{ZM93} that supports fast random access, relative Lempel-Ziv~\cite{KPZ10} (RLZ) is one of the most theoretically interesting of those variants.

The LZ77 parse of a text $T [1..n]$ is a partition of $T$ into $z$ phrases such that, for each phrase $T [i..j]$, the phrase itself does not occur in $T [1..j - 1]$ but its (possibly empty) longest proper prefix $T [i..j - 1]$ does.  (For simplicity, it is common to assume that $T [n]$ is a unique end-of-file symbol, so the last phrase also fits this definition.)  We store each LZ77 phrase as a triple indicating that phrase's length minus 1, the position in $T$ of the leftmost occurrence of its longest proper prefix, and its last character.  The LZ77 parse can be built in $O (n)$ time by building first a suffix tree for $T$, but this takes $O (n)$ workspace as well.  In the past few years researchers have shown how we can build the LZ77 parse with one pass over $T$ and slightly super-linear time and compressed workspace, by building incrementally a compressed self-index of $T$ while simultaneously using that index to parse $T$~\cite{PP18,NIIBT20}.

If $T$ is a concatenation of genomes from many individuals of the same or closely-related species, then it is natural to store the first genome or first few genomes uncompressed as a reference with which to compress the other genomes.  This is the main idea behind RLZ: given $T$ and $\ell$, we store $R = T [1..\ell]$ uncompressed as a reference and greedily parse $T [\ell + 1..n]$ into $z'$ phrases that each occur in $R$.  (For simplicity, it is common to assume the reference contains all the characters in the alphabet.)  We store each RLZ phrase as a pair indicating that phrase's length and the position in $R$ of its leftmost occurrence.  If we start by building an FM-index for $R$, then we can build the RLZ parse of $T [\ell + 1..n]$ with respect to $R$ with one pass over $T$ and slightly super-linear time (or even linear time if the alphabet is polylogarithm in $n$) and $O (\ell + z')$ workspace.

Kosolobov et al.~\cite{KVNP20} proposed computing the LZ77 parse of the RLZ parse and then converting it into an LZ77-like parse of $T$, but this produces only an approximation of the LZ77 parse of $T$.  In this paper we show how, given $R$ and the RLZ parse of $T [\ell + 1..n]$ with respect to $R$, we can build the exact LZ77 parse of $T$ in $n\,\polylog (n)$ time and $O (\ell + z')$ total space.

\section{Data structures}

Suppose we are given $R = T [1..\ell]$ and the RLZ parse of $T [\ell + 1..n]$ with respect to $R$, which together take $O (\ell + z')$ space.  Farach and Thorup~\cite{FT98} observed that, by the definition of LZ77, the first occurrence of any substring in $T$ touches an LZ77 phrase boundary.  Although this is not generally true of the RLZ parse of $T [\ell + 1..n]$ with respect to $R$, it is if we build the LZ77 parse of $R$ and consider its phrases as the leading phrases of the RLZ parse of $T [\ell + 1..n]$ with respect to $R$.  For brevity, from now on we will refer to this combined parse simply as the RLZ parse (of $T$).  Notice it has $O (\ell + z')$ phrases and we can build and store it in $(\ell + z')\,\polylog (n)$ time and $O (\ell + z')$ space.

To support $\polylog (n)$-time random access to $T$, we store in an $O (\ell + z')$-space predecessor data structure the starting position $i$ in $T$ of each RLZ phrase $T [i..j]$, with the starting position $i'$ in $R$ of that phrase's leftmost occurrence as satellite data.  Given $k \leq n$, if $k \leq \ell$ then we can return $T [k] = R [k]$ immediately; otherwise, we find the largest starting position $i \leq k$ of an RLZ phrase, look up the starting position $i'$ in $R$ of that phrase's leftmost occurrence, and return $T [k] = R [i' + k - i]$ in $\polylog (n)$ time.

It takes $O (\ell)$ time to compute and store the Karp-Rabin hash of each prefix of $R$, and then it takes $O (z'\,\polylog (n))$ time to compute and store the hash of each prefix of $T$ ending at an RLZ phrase boundary.  Storing all these hashes takes $O (\ell + z')$ space and allows us to compute the hash of any substring of $T$ in $\polylog (n)$ time (by breaking it into the suffix of an RLZ phrase, a sequence of complete RLZ phrases, and a prefix of an RLZ phrase, and combining the appropriate hashes).  Once we can do this, we can find the length of the longest common prefix of suffixes of $T$ in $\polylog (n)$ time, using binary search and checking substring equality by comparing hashes; we can then lexicographically or co-lexicographically compare substrings of $T$ in $\polylog (n)$ time, by checking the first characters after the longest common prefix; so we can co-lexicographically sort the RLZ phrases and lexicographically sort the suffixes of $T$ starting at RLZ phrase boundaries, all in $(\ell + z')\,\polylog (n) \in n\,\polylog (n)$ time with high probability.  (With Karp-Rabin hashing our results are Monte-Carlo randomized, but with more sophisticated techniques~\cite{GKKLS18} we can make them Las-Vegas randomized.)

In $(\ell + z')\,\polylog (n)$ time we build an $(\ell + z') \times (\ell + z')$ grid on which there is a point at $(x, y)$ with weight $w$ if the co-lexicographically $x$th RLZ phrase ends at $w$ and is immediately followed by the lexicographically $y$th suffix of $T$ starting at an RLZ phrase boundary.  We store this grid in $O (\ell + z')$ space such that it supports $\polylog (n)$-time 2-dimensional range-minimum queries~\cite{Nav14}.  Given a pattern $P [1..m]$ split into $P [1..i]$ and $P [i + 1..m]$ and the Karp-Rabin hashes of $P [1..i]$ and $P [i + 1..m]$, in $\polylog (n)$ time we can find the co-lexicographic range of RLZ phrases ending with $P [1..i]$ and the lexicographic range of suffixes of $T$ starting with $P [i + 1..m]$ at RLZ phrase boundaries, and then report the leftmost starting position of an occurrence of $P$ in $T$ that is split by an RLZ phrase boundary into $P [1..i]$ and $P [i + 1..m]$ (if such an occurrence exists).

\begin{lemma}
Given $R = T [1..\ell]$ and the RLZ parse of $T [\ell + 1..n]$ with respect to $R$, in $(\ell + z')\,\polylog (n)$ time and $O (\ell + z')$ workspace we can build an $O (\ell + z')$-space index with which, given a pattern $P [1..m]$ split into $P [1..i]$ and $P [i + 1..m]$ and the Karp-Rabin hashes of $P [1..i]$ and $P [i + 1..m]$, in $\polylog (n)$ time we can report the leftmost starting position of an occurrence of $P$ in $T$ that is split by an RLZ phrase boundary into $P [1..i]$ and $P [i + 1..m]$ (if such an occurrence exists).
\end{lemma}

\pagebreak

We note as an aside that, although our index is static, our use of a range-minimum data structure effectively endows it with a kind of partial persistence: if we want to know if an occurrence of $P [1..m]$ starting in $T [1..j]$ is split by an RLZ phrase boundary into $P [1..i]$ and $P [i + 1..m]$, then we can query our index and ignore the answer if it is larger than $j$.  In this way, we are following the approach of researchers who build the LZ77 parse by building compressed self-indexes incrementally.

\section{Algorithm}

To see how we build the LZ77 parse of $P$ with our index, suppose we have already computed some number of LZ77 phrases and the next LZ77 phrase is $T [i + 1..k]$, although we do not yet know $k$.  This means that, for $k' < k$, the substring $T [i + 1..k']$ occurs in $T [1..k' - 1]$ --- so there is some way to split $T [i + 1..k']$ into $T [i + 1..j]$ and $T [j + 1..k']$ such that an occurrence of $T [i + 1..k']$ in $T [1..k' - 1]$ is split by an RLZ phrase boundary into $T [i + 1..j]$ and $T [j + 1..k']$.  (We allow the possibility that $k_j = j$, so $T [j + 1..k']$ is empty.)  On the other hand, for $k' \geq k$, the substring $T [i + 1..k']$ does not occur in $T [1..k' - 1]$.

Our idea is to use our index and binary search to find, for each value $j$ from $i + 1$ to $k - 1$ in turn, the largest $k_j$ such that $T [1..k_j]$ occurs in $T [1..k_j - 1]$ split by an RLZ phrase boundary into $T [i + 1..j]$ and $T [j + 1..k_j]$ and, if such a $k_j$ exists, the starting position of the leftmost such occurrence in $T$ of $T [i + 1..k _j]$.  This takes $\polylog (n)$ time for each value of $j$.  We do not store all these $k_j$ values and starting positions, since that might take more than $O (\ell + z')$ space; instead, we store only the largest $k_j$ value we have seen so far, which we call $k_{\max}$, and the starting position of the leftmost occurrence of $T [i + 1..k_{\max}]$ in $T [1..k_{\max} - 1]$ that we have found so far, which we call $s_{\max}$.

Suppose that, first, when we query our index to find the largest $k_j$ such that $T [1..k_j]$ occurs in $T [1..k_j - 1]$ split by an RLZ phrase boundary into $T [i + 1..j]$ and $T [j + 1..k_j]$, we learn there is no occurrence of $T [i + 1..j]$ immediately preceding an RLZ phrase boundary; second, when we reach this point, $k_{\max} = j - 1$.  It follows that $j = k$, so we can report the triple $(k_{\max} - i, s_{\max}, T [k_{\max} + 1])$ that encodes the next phrase.  Figure~\ref{fig:code} shows pseudo-code for this procedure.  We spend a total of $(k - i)\,\polylog (n)$ finding the triple that encodes the next phrase, which has length $k - i$, so we use $n\,\polylog (n)$ total time and $O (\ell + z')$ space building the LZ77 parse.

\begin{figure}[t]
\begin{verbatim}
kmax = i
smax = n + 1
for j from i + 1 to n    
    let kj be the largest value such that T[i + 1..kj] occurs in T[1..kj - 1]
      split by an RLZ phrase boundary into T[i + 1..j] and T[j + 1..kj]    
    let sj be the leftmost starting position in T of such an occurrence of
      T [i + 1..kj]    
    if kj > kmax then
        kmax = kj
        smax = sj
    end if    
    if kj == kmax and smax > sj then
        smax = sj
    end if    
    if kmax == j - 1 then
        break
    end if    
end for
return (kmax - i, smax, T[kmax + 1])
\end{verbatim}
\caption{Pseudo-code for computing an LZ77 phrase starting at $T [i + 1]$, in time $\polylog (n)$ times the length of the phrase and $O (\ell + z')$ space.}
\label{fig:code}
\end{figure}

\begin{theorem}
Consider a text $T [1..n]$ prefixed by a reference sequence $R = T [1..\ell]$.  Given $R$ and the $z'$-phrase relative Lempel-Ziv parse of $T [\ell + 1..n]$ with respect to $R$, we can build the LZ77 parse of $T$ in $n\,\polylog (n)$ time and $O (\ell + z')$ total space.
\end{theorem}

\end{document}